\documentclass[%
aip,
amsmath,amssymb,
preprint,%
]{revtex4-1}

\usepackage{graphicx}
\usepackage{dcolumn}
\usepackage{bm}
\usepackage{xcolor}
\usepackage{booktabs}
\usepackage[utf8]{inputenc}
\usepackage[T1]{fontenc}
\usepackage{mathptmx}
\usepackage{etoolbox}
\usepackage{hyperref}

\makeatletter
\def\@email#1#2{%
	\endgroup
	\patchcmd{\titleblock@produce}
	{\frontmatter@RRAPformat}
	{\frontmatter@RRAPformat{\produce@RRAP{*#1\href{mailto:#2}{#2}}}\frontmatter@RRAPformat}
	{}{}
}%
\makeatother
\begin{document}
	
	
	\title[]{Modeling inductive radio frequency coupling in powerful negative hydrogen ion sources: optimizing the RF coupling}
	
	\author{D. Zielke}
	\author{S. Briefi}
	
	\affiliation{Max-Planck-Institut für Plasmaphysik, Boltzmannstr. 2, 85748 Garching, Germany}
	
	\author{U. Fantz}
	\affiliation{Max-Planck-Institut für Plasmaphysik, Boltzmannstr. 2, 85748 Garching, Germany}
	\affiliation{AG Experimentelle Plasmaphysik, Universität Augsburg, 86135 Augsburg, Germany}
	
	\email{dominikus.zielke@ipp.mpg.de}
	
	\date{\today}
	
	\begin{abstract}
		In the fusion experiment ITER powerful neutral beam injection (NBI) systems will be used. The NBI's core component is a negative hydrogen ion source, which is based on a modular concept. Eight cylindrical drivers, each having a volume of several liters, are attached to one common expansion and extraction region. Within the drivers an inductively coupled plasma is sustained by an external cylindrical coil at filling pressures not larger than 0.3\,Pa. Radio frequency (RF) generators operating at a driving frequency of 1\,MHz feed the coils via a matching network with powers of up to 100\,kW per driver. These high powers entail high voltages, which make the ion source prone to electrical breakdowns and arcing, wherefore its reliability is reduced. Moreover, at the ITER prototype RF ion source not more than 60\% of the power is absorbed by the plasma, whereas the rest is lost for heating the RF coil and conducting structures of the driver. To optimize the power coupling in the prototype source, a previously validated self-consistent fluid-electromagnetic model is applied. The optimization studies reveal a complex interplay between network losses (mainly caused by the skin effect and eddy currents), and nonlinear plasma phenomena, such as the RF Lorentz force. The model demonstrates promising optimization concepts for the RF coupling in future NBI ion sources. In particular, by increasing the axial driver length and the driving frequency it is possible to enhance the fraction of absorbed power to values around 90\%.
	\end{abstract}
	
	\maketitle

	\section{\label{sec:Introduction}Introduction}
	
\begin{figure}[h!]
		\centering
		\includegraphics[width=0.5\textwidth]{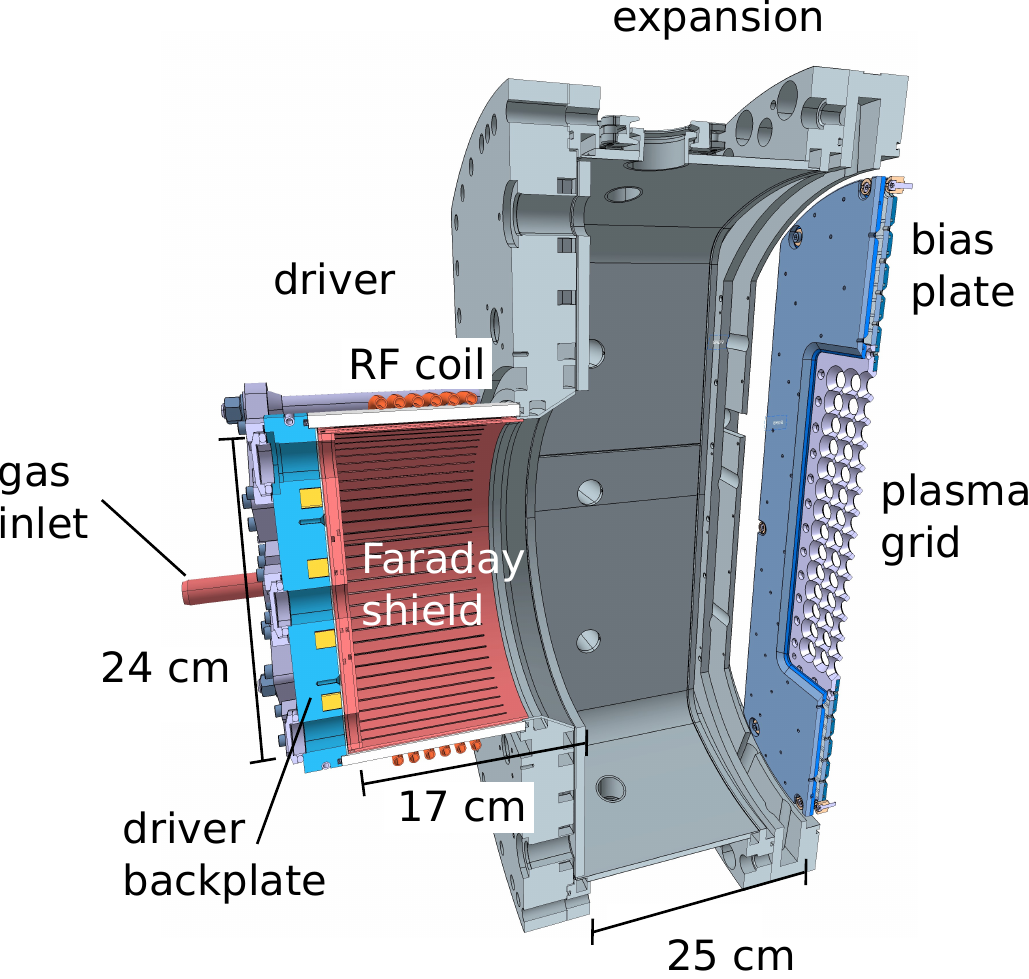}
		\caption{Schematic of the ITER prototype RF ion source at the Batman Upgrade test bed. Figure taken from Zielke et al..\cite{Zielke_2022}}
		\label{fig:BUGSetup}
	\end{figure}

	Large and powerful systems for neutral beam injection (NBI) will be used in the ITER fusion experiment.\cite{ITER} The main goal for the ITER NBI is to produce a beam of highly energetic neutral deuterium atoms for one hour of continuous operation to heat the fusion plasma.\cite{Hemsworth_2017} The NBI consists of multiple components, each being designed and optimized for a specific purpose. One of the NBI's core components is the negative ion source. Figure~\ref{fig:BUGSetup} shows a cutaway drawing of the ITER prototype RF ion source at the Batman Upgrade test bed,\cite{Heinemann_2017} which is 1/8 of the full ITER ion source size.
	
	The ion source is composed of a modular cylindrical driver, which is attached to a rectangular expansion. Plasma is produced in the driver and flows into the expansion, where negative ions are extracted at the plasma grid. To achieve the main goal for the NBI as introduced above, it is foreseen that a large number of several 100\,A\,m$^{-2}$ of negative hydrogen ions have to be produced and extracted for a duration of up to one hour without the ion source to fail or break down. That means the ion source has to operate in a highly reliable way.
	
	The majority of negative ions are produced on the surface of the plasma grid. In this process one or two surface electrons are attached to an impinging neutral atom or positive ion, respectively. There are several measures to enhance the surface production of negative ions. A fundamental one is to produce a dense and hot plasma in the driver. Here typical electron densities and temperatures are in the order of $10^{18}\,$m$^{-3}$ and 10\,eV, respectively. The large plasma density has two positive effects on the production of negative ions. Firstly, it directly results in large ion fluxes towards the expansion and onto the plasma grid. Secondly, the high dissociation degree, which increases with plasma density and electron temperature, ensures large atomic fluxes onto the plasma grid. Another measure which is routinely applied is evaporating cesium into the ion source, because this lowers the work function of the plasma grid material and increases the probability for the electrons to leave the surface and get attached to an impinging atom or positive ion.
	
	The positive extraction voltage at the plasma grid not only attracts negative ions but also electrons. The latter are co-extracted and bent by magnets onto the second grid behind the plasma grid, where they produce large heat loads. Hence the performance of the ion source is limited by these electrons. To decrease the amount of co-extracted electrons a horizontal magnetic filter field is applied in the expansion chamber. In ITER sources, this will be done by a direct current in the kA range, which flows vertically through the plasma grid. The resulting filter field strength is in the range of a few tens of Gauss in close vicinity to the plasma grid and decreases towards the driver, where it is below 10\,G. Electron transport across the magnetic filter field is decreased. Therefore electrons coming from the driver are rarefied to typically around $10^{17}\,$m$^{-3}$ after having traveled through the filter, where an increased number of collisions act to cool down the electron temperature to a value around 1\,eV. Applying a potential at the plasma grid and/or the bias plate (see figure~\ref{fig:BUGSetup}, right) decreases the number of co-extracted electrons even further. However, filter field as well as potentials lead to a drift motion of the plasma, wherefore unwanted vertical asymmetries appear.
	
	To produce the dense and hot driver plasma, hydrogen gas is let into the driver, such that a maximum filling pressure (without discharge) of 0.3\,Pa is reached in the ion source. At this low pressure the losses of negative ions in the extraction and acceleration grid system via collisions with neutrals are reduced to an acceptable level of 30\%.\cite{Krylov_2006} An inductively coupled plasma is sustained within the driver via an external cylindrical coil, which has six windings at the ion source of Batman Upgrade. The coil is connected to an RF power generator via a capacitive matching network. Up to 100\,kW of power is typically fed into the driver at a driving frequency of 1\,MHz. The large powers at the low applied frequency are related to high voltages at the RF coil and in the matching network. As a consequence, the system is prone to electrical breakdowns and arcs, wherefore special measures are necessary\cite{Fantz_2017} and the ion source's reliability is decreased.
	
	Not all of the power delivered by the RF generator is coupled to the plasma but lost for heating - mainly of the RF coil and Faraday shield,\cite{Zielke_2021} where eddy currents are driven in a skin layer, whose thickness is in the order of 10\,$\mu$m. The internal Faraday shield, as depicted in figure~\ref{fig:BUGSetup}, is necessary to protect the dielectric cylindrical driver walls against plasma erosion. The large losses in the Faraday shield and coil are one of the reasons (besides the plasma load onto the Faraday shield) why active water cooling of these components is necessary. Furthermore, the matching transformer, which is part of the matching circuit at the Batman Upgrade ion source, is subject to non-negligible magnetization losses.\cite{Zielke_2021}
	
	The ratio of power absorbed by the plasma and total delivered power by the generator is called RF power transfer efficiency $\eta$. It is an important quantity to assess how many losses there are and how much power can be coupled to the plasma. The latter is directly related to the plasma parameters (especially to the electron density), whereas the generator power only quantifies how much active power is delivered by the RF generator. In the ion sources at Batman Upgrade as well as in the one of SPIDER, which is the full ITER size ion source,\cite{Toigo_2021} measurements were performed. Using no matching transformer in both cases, $\eta$ was recently determined to range only between 50\% and 60\%, depending on the operation conditions.\cite{Zielke_2021, Jain_2022} Hence there is considerable optimization potential. By maximizing $\eta$, it would become possible to produce a dense and hot plasma in the driver, while using lower generator powers and voltages. Hence, the reliability of ion source operation increases considerably.
	
	The envisaged optimization of the RF coupling is non-trivial, since there are various external parameters, such as driving frequency, power, feed gas type (i.e.\ hydrogen or deuterium), filling pressure as well as coil and discharge geometry. Additionally, the coil voltage and $\eta$ also depend on the presence of magnetic fields. In the case of the RF negative ion source there is the magnetic filter field, which is non-negligible in the driver, as introduced above. And a cusp field, formed by permanent magnets in the driver back plate, as indicated in yellow in figure~\ref{fig:BUGSetup}. Its purpose is to reduce plasma losses at the driver back plate. All these parameters affect the losses in the network components (i.e.\ RF coil and Faraday shield) as well as the power coupling to the plasma in a complicated way.
	
	A modeling approach where the RF power coupling is described self-consistently can capture how the plasma and the electromagnetic fields are affected by the external parameters. However, the complexity of the nonlinear coupling mechanism in the low pressure, high power, low frequency regime imposes huge modeling challenges.\cite{Zielke_2022, Hagelaar_2011} Previous attempts to describe the RF power coupling were not self-consistent and important nonlinear effects were missing.\cite{Hagelaar_2011, Jain_2018_1, Chen_2021} Also there was no possibility to validate previous models, since experimental measurements of the RF power transfer efficiency in large and powerful ion sources became available only recently.\cite{Zielke_2021} Using this comprehensive data set it was possible to establish and validate a self-consistent fluid-electromagnetic model.\cite{Zielke_2022, Zielke_thesis_2021}
	
	In particular it was shown in the model validation that the power coupling in RF ion sources relies on a complex nonlinear interplay between the RF Lorentz force and RF current diffusion. Due to the low driving frequency of 1\,MHz there is a large magnetic RF field of more than 100\,G in the driver. Hence the Lorentz force acts to push the plasma away from the RF coil. This effect is mitigated by diffusion of the RF current. Another consequence of the large magnetic RF field is that the driver plasma is in the local RF skin effect regime, wherefore a local approximation of the viscosity produces the amount of RF current diffusion, which yields calculated results in reasonable agreement with experimentally obtained electrical and plasma quantities.\cite{Zielke_2022} Beyond that it was revealed that depletion of the neutrals via ionization has a profound impact on the RF power coupling, the connecting quantity being again the electron viscosity, which increases nonlinearly when the neutral density decreases. In this way it was possible for the first time to model ion source discharges at fusion relevant powers of up to 100\,kW per driver. With these gained insights it becomes now possible to apply the self-consistent model to guide the experimental efforts towards an optimized RF power coupling.
	
	In section~\ref{sec:ModelingApproach} of this work the modeling approach is outlined and the figures of merit, which serve as a basis for the optimization of the RF power coupling, are defined. This is followed by optimization studies in sections~\ref{sec:OptimizingRFCoil},~\ref{sec:OptimizingDischargeGeometry} and~\ref{sec:OptimizingRF}, where the governing physics effects are analyzed, when the RF coil, discharge geometry, and applied frequency are changed. The most  beneficial measures for the power coupling are combined in section~\ref{sec:GlobalOptimum}. In the final section~\ref{sec:Conclusion}, concluding remarks and an outline for necessary directions of further model development are given.

\section{Self-consistent modeling approach and power coupling figures of merit}
\label{sec:ModelingApproach}

The modeling approach is described using the ITER RF prototype negative ion source at the Batman Upgrade test bed. Two separate models are combined for the self-consistent description of the RF power coupling.

First, a source impedance model is used to calculate the network resistance and inductance. The simulation domain is based on the 3D CAD model, as shown in figure~\ref{fig:BUGSetup}. The expansion region as well as peripheral support structure such as screws and bolts have been excluded from the simulation domain, since their contribution to the network impedance is found to be negligible.\cite{Briefi_2023} The same is true for the matching capacitors as well as the transmission line, which connects the RF generator to the matching network. The matching transformer, which was found to produce additional losses at the Batman Upgrade ion source,\cite{Zielke_2021} is not considered in this work, since it is not foreseen at ITER. A vacuum is assumed within the driver, wherefore plasma is not simulated in the source impedance model.

\begin{figure}[h!]
	\centering
	\includegraphics[width=0.8\textwidth]{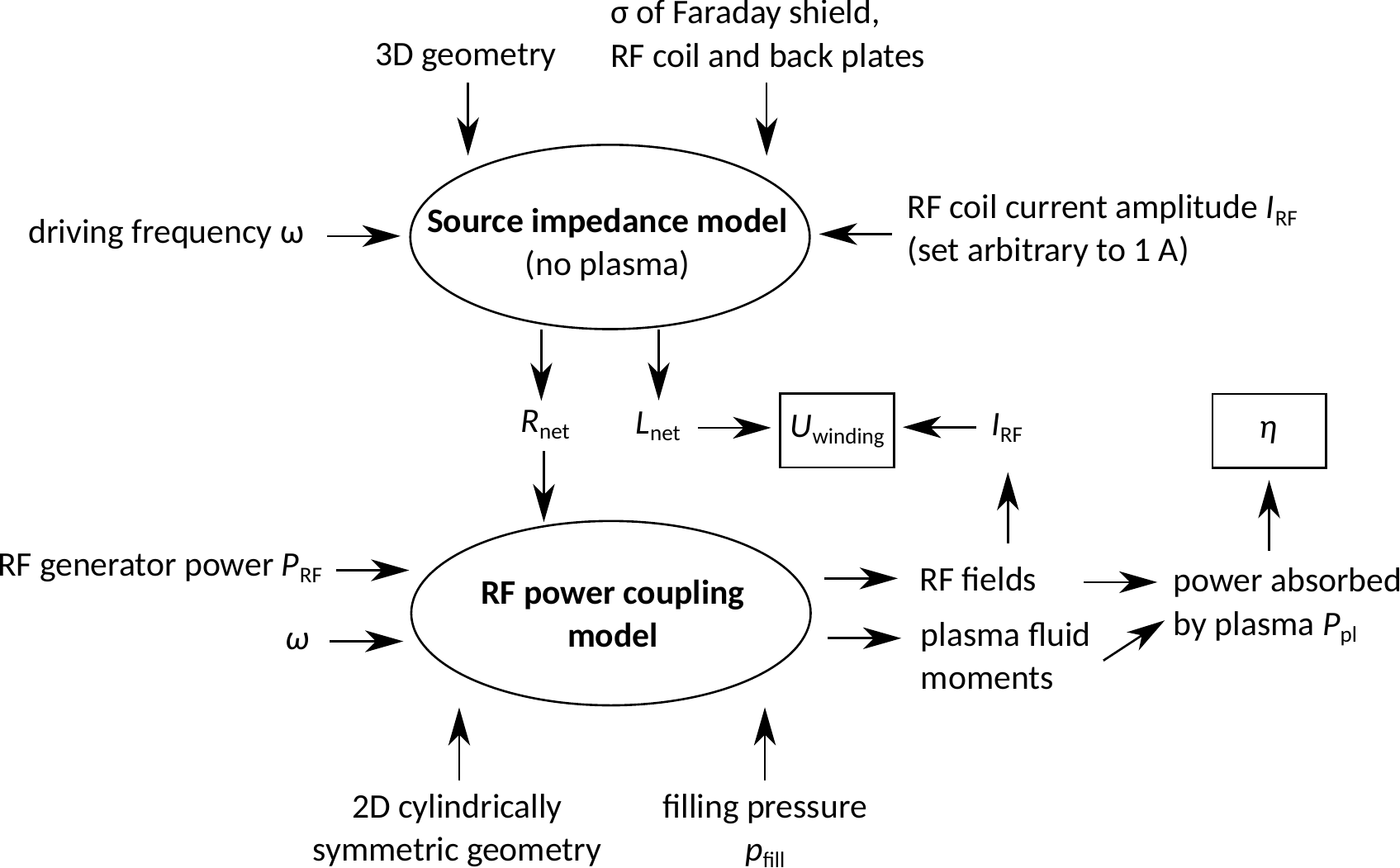}
	\caption{Flow diagram showing the relation between the inputs and outputs of the source impedance model and the RF power coupling model.}
	\label{fig:ModelWorkflow}
\end{figure}

Maxwell's equations are solved to calculate the network resistance and inductance. The former results from Joule heating losses in the six windings of the RF coil, eddy current losses in the Faraday shield and in the driver- and source back plates. Because of the linearity of Maxwell's equations, an arbitrary RF current amplitude of $I_\mathrm{RF} = 1$\,A is used at the coil as input to excite the system. The numerical effort is considerably reduced by choosing a time-harmonic formulation. This enables using boundary conditions on all conducting surfaces (having electrical conductivities $\sigma$ as input values),\cite{Briefi_2023} instead of resolving the inside, where a small skin layer of 65\,$\mu$m at 1\,MHz in copper is numerically prohibitive. Because of the absence of plasma, no nonlinear effects occur, wherefore the time-harmonic approximation is well justified. The output of the model is the complex RF network impedance $Z_\mathrm{net} = R_\mathrm{net} + \mathrm{i} \omega L_\mathrm{net}$, where $R_\mathrm{net}$, $L_\mathrm{net}$ and i denote the network resistance, inductance and the imaginary unit, respectively. The angular driving frequency $\omega = 2 \pi f$. The model flow diagram~\ref{fig:ModelWorkflow} summarizes at the top the input and output quantities from the source impedance model.

The source impedance model is implemented in the AC/DC module of Comsol Multiphysics\textsuperscript{\textregistered}.\cite{comsol_60} The model runtime is in the order of one hour on a six-core workstation (Intel64 Family 6 Model 62 processors running at 3.5\,GHz) with 128\,GB of RAM. The model was successfully validated by comparing the calculated impedance to the one that was experimentally determined at the Batman Upgrade test bed. Agreement within 20\% for the network resistance and within 2\% for the network inductance was obtained. Briefi et al.\cite{Briefi_2023} give more detailed information regarding the validation of the source impedance model. 

\begin{figure}[h!]
	\centering
	\includegraphics[width=0.5\textwidth]{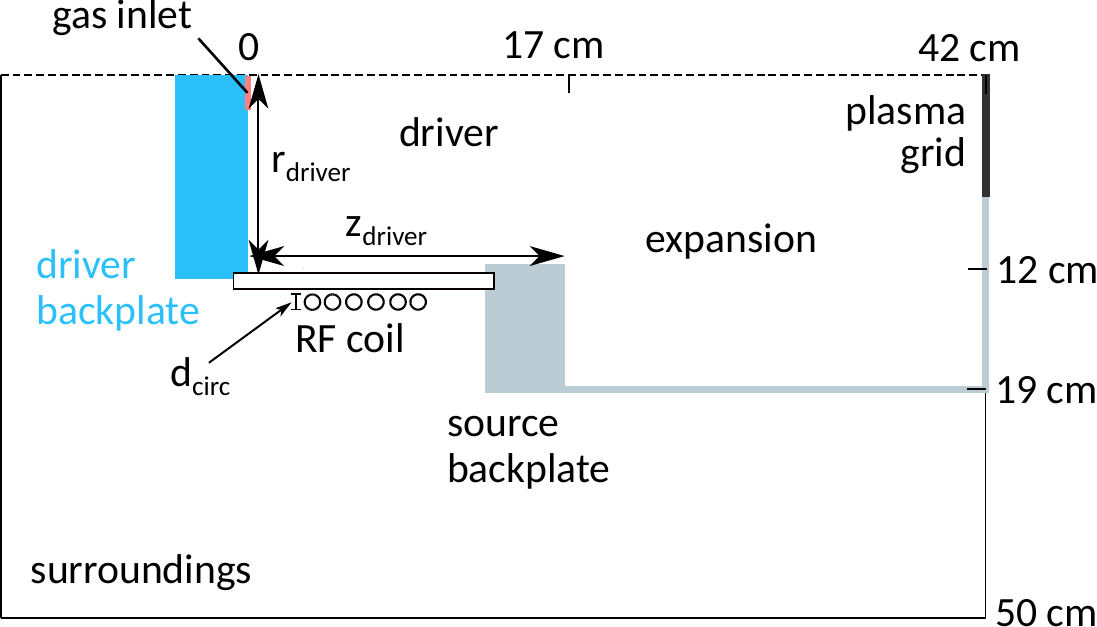}
	\caption{2D cylindrically symmetric simulation domain of the RF power coupling model. Figure taken from Zielke et al..\cite{Zielke_2022}}
	\label{fig:SimulationDomain}
\end{figure}

For the self-consistent description of the power coupling between RF fields and plasma, a 2D cylindrically symmetric model is used.\cite{Zielke_2022} The simulation domain of this model is shown in figure~\ref{fig:SimulationDomain}.

The fluid approximation is employed to simulate the various different neutral and plasma species. Hydrogen is used as working gas. It was found experimentally that in deuterium the RF power transfer efficiency is increased by around 5\%, when compared to hydrogen.\cite{Zielke_2021} This effect coincides with a slightly increased electron density. Since the effect is present at different pressures and powers, it is expected that all trends obtained in hydrogen should hold for deuterium as well. An initial run with neutral gas is performed, where Navier-Stokes particle, momentum and energy balance are solved in the driver and expansion region for hydrogen molecules only. As boundary conditions, the experimentally measured inflow is set at the gas inlet (see figure~\ref{fig:SimulationDomain}) and the transparency of the plasma grid adjusted, such that the experimentally measured filling pressure of typically 0.3\,Pa is reproduced by the model. No-slip boundary conditions are set at all other walls. After this initial step the model is run again using the same inlet flux of molecular hydrogen and grid transparency. However, at this point also separate Navier-Stokes equation (i.e.\ particle, momentum and energy balance) are solved for the atoms. The transport of positive ions H$^+$, H$_2^+$ and H$_3^+$ is modeled by separate particle and momentum balances, where the important advection term is retained. No energy balances are solved for the ions, since it is assumed that the energy exchange between ions and neutrals is very efficient due to the similar masses. This leads to the assumptions that  $T_{\mathrm{H}^+} \approx T_\mathrm{H}$ and $T_{\mathrm{H}_2^+} \approx T_{\mathrm{H}_3^+} \approx T_{\mathrm{H}_2}$.

The core element of the model is the self-consistent description of the coupling between the electrons and the RF fields. The electrons are described by a particle, momentum and energy balance, and the RF fields by Maxwells equations. A magnetic RF field is excited by applying a surface current amplitude $J_\mathrm{surf} = I_\mathrm{RF}/(\pi d_\mathrm{circ})$ at each RF coil winding, where $d_\mathrm{circ}$ is the diameter of each  winding (see figure~\ref{fig:SimulationDomain}). The magnitude of $J_\mathrm{surf}$ is controlled by an integral controller, which acts to fulfill the power balance $P_\mathrm{RF} = P_\mathrm{net} + P_\mathrm{pl}$ at every instant in time. Hereby the root mean square value of the generator output active power $P_\mathrm{RF}$ is a set input value, and the power lost in the network $P_\mathrm{net} = \frac{1}{2}R_\mathrm{net}I_\mathrm{RF}^2$ is calculated using the network resistance as obtained from the source impedance model. The power absorbed by the plasma $P_\mathrm{pl}$ is calculated from a volume integral over the power density in the cylindrically symmetric discharge volume. The power density in turn is the product of the RF current density, which results from the momentum balance of electrons, and the induced electric RF field, which results from Faraday's law. Because of the low frequency and the high power density, the magnetic RF field (calculated from Amp\`{e}re's law) is rather large, wherefore the discharge is in the transition region between the nonlinear and the local skin effect regime.\cite{Froese_2009} It was found by Zielke et al.\cite{Zielke_2022} that in this regime the RF Lorentz force and a collisional (Navier-Stokes) approximation for the viscosity have to be retained, to reach good agreement with the experimentally measured RF coil currents and plasma parameters. The validation of the RF power coupling model is described in detail in a dedicated paper.\cite{Zielke_2022} Herein power and pressures variations are performed and calculated electrical quantities such as the RF coil current as well as local plasma parameters electron density and temperature are compared to their experimentally obtained counterparts. Overall good agreement between measurements and model results is obtained. An overview over the inputs and outputs of this model is depicted in the bottom of figure~\ref{fig:ModelWorkflow}.

The two important figures of merit for the optimization of the RF power coupling are the RF power transfer efficiency $\eta$ and the voltage per coil winding $U_\mathrm{winding}$. Since only a negligible part of the RF power is radiated away, $\eta$ is calculated as
\begin{equation}
	\eta = \frac{P_\mathrm{pl}}{P_\mathrm{pl} + P_\mathrm{net}} = \frac{R_\mathrm{pl}}{R_\mathrm{pl} + R_\mathrm{net}},
	\label{eq:eta}
\end{equation}
where the root mean square powers $P_\mathrm{j} = \frac{1}{2}R_\mathrm{j} I_\mathrm{RF}^2, \mathrm{j}\in \{\mathrm{pl}, \mathrm{net}\}$ and the RF coil current amplitude is denoted as $I_\mathrm{RF}$. $R_\mathrm{net}$ on the right-hand-side of equation~(\ref{eq:eta}) is calculated in the so called source impedance model, which is 3D. The plasma equivalent resistance $R_\mathrm{pl}$ is obtained from the 2D RF power coupling model via the volume integral over the calculated power absorption profile (which yields $P_\mathrm{pl}$) and the applied RF coil current amplitude $I_\mathrm{RF}$, which is also calculated self-consistently (see figure~\ref{fig:ModelWorkflow}). Since typically $R_\mathrm{net} \ll \omega L_\mathrm{net}$, the voltage per coil winding can be approximated as 
\begin{equation}
	U_\mathrm{winding} \approx \frac{\omega L_\mathrm{net}I_\mathrm{RF}}{N_\mathrm{windings}}.
	\label{eq:U_winding}
\end{equation}
Herein, the network inductance $L_\mathrm{net}$ is calculated in the source impedance model, whereas $I_\mathrm{RF}$ is resulting from the RF power coupling model. The number of coil windings are denoted by $N_\mathrm{windings}$. In the ITER NBI prototype ion source at the Batman Upgrade test bed $N_\mathrm{windings} = 6$.

In the following, several external parameters are investigated. Only one parameter at a time is varied, as this allows to study how this parameter affects the physics of the RF coupling. The investigated parameters are the coil geometry, the discharge geometry, and the driving frequency. As for the optimization of the RF coil, $N_\mathrm{windings}$ is investigated as external parameter. Regarding the geometry, the driver radius $r_\mathrm{driver}$ and axial length $z_\mathrm{driver}$, as shown in figure~\ref{fig:SimulationDomain}, are varied. The expansion geometry is left unchanged during the optimization studies, since it does not affect the RF power coupling that takes place in the driver. In the following optimization studies the filling pressure is fixed at the required ITER pressure of $p_\mathrm{fill} = 0.3\,$Pa. The total delivered power by the RF generator is set to a typical value of $P_\mathrm{RF} = 60\,$kW. Note that $P_\mathrm{RF}$ and not $P_\mathrm{pl}$ is set. This is done to ease comparability to the experiment. Table~\ref{tab:Parameters} summarizes the most important model input parameters and their values as a baseline for the optimization studies.

\begin{table}[h!]
	\caption{Parameters of the ion source at the Batman Upgrade test bed as a baseline for the optimization.}
	\label{tab:Parameters}
	\centering
	\small{
		\begin{ruledtabular}
		\begin{tabular}{ccccc}
			Parameter name & Parameter value  \\ 
			\hline
			Generator power & $P_\mathrm{RF} = 60\,\mathrm{kW}$ \\
			Filling pressure & $p_\mathrm{fill} = 0.3\,\mathrm{Pa}$ \\
			Driving frequency & $f = \omega/2\pi =1\,$MHz \\
			Number RF coil windings & $N_\mathrm{windings} = 6$ \\
			Driver axial length & $z_\mathrm{driver} = 17.4\,$cm  \\
			Driver radius & $r_\mathrm{driver} = 11.85\,$cm \\
		\end{tabular}
		\end{ruledtabular}
	}
\end{table}
In the 2D cylindrically symmetric model formulation it is not possible to include the 3D topologies of the two external magnetic fields, i.e.\ the magnetic filter field and cusp field in the driver backplate, which are present in the ion source at the Batman Upgrade test bed. However, since the magnetic filter field is dominant only in the expansion, it is expected that it does not influence $\eta$ significantly. This has been observed experimentally by Zielke et al.\cite{Zielke_2021} for the single-driver ion source at the Batman Upgrade test bed. Note that it was found by Jain et al.~that the filter field has an effect on the RF power coupling in multi-driver ion sources.\cite{Jain_2022}

The highly non-uniform magnetic field produced by the cusp magnets in the driver back plate decreases axially very steeply into the driver. As discussed by Zielke,\cite{Zielke_thesis_2021} using the steep decrease in the cylindrically symmetric RF power coupling model results in a decreased RF power transfer efficiency. This is because the magnetic field acts to form an axial transport barrier, which pushes the electrons further into the inside of the driver. Hence the region, where the RF fields can couple to the electrons is made smaller.

\section{Optimizing the RF coil}
\label{sec:OptimizingRFCoil}

\begin{figure}[h!]
	\centering
	\includegraphics[width=0.9\textwidth]{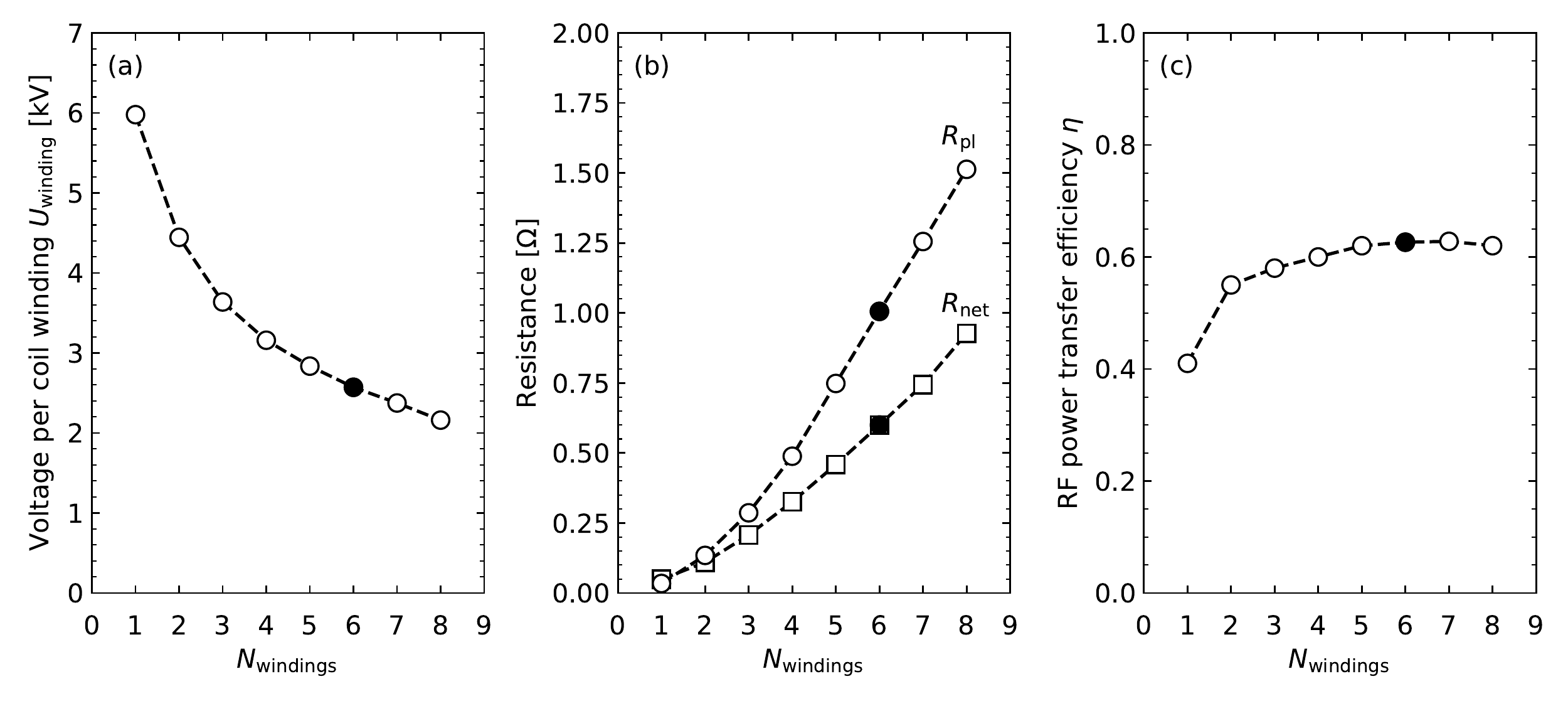}
	\caption{(a) Voltage per coil winding $U_\mathrm{winding}$, (b) equivalent plasma resistance $R_\mathrm{pl}$, network resistance $R_\mathrm{net}$ and (c) RF power transfer efficiency $\eta$ as a function of the number of coil windings $N_\mathrm{windings}$. The full symbols indicate the validated values of the present experimental setup at $N_\mathrm{windings} = 6$. Figures (a) and (b) are adapted from Zielke,\cite{Zielke_thesis_2021} whereas (c) is reproduced from Briefi et al..\cite{Briefi_2022}}
	\label{fig:NumberOfCoilWindings}
\end{figure}

Investigations of the axial position of the RF coil revealed that the coil windings should be spread around the axial driver center, as indicated in figure~\ref{fig:SimulationDomain}. In this case the electric field distribution reaches as much plasma volume as possible and at the same time the distance of the outermost coil winding to the backplates is at a maximum avoiding a pinched electric field.\cite{Zielke_thesis_2021} Using this axially central position the number of coil windings is varied between one and eight. The corresponding modeling results are shown in figure~\ref{fig:NumberOfCoilWindings}.

The inductance of the RF coil is larger when windings are added to the coil, wherefore also $L_\mathrm{net}(N_\mathrm{windings})$ is a monotonically increasing function. The larger inductance goes along with a lower RF coil current, wherefore the corresponding $I_\mathrm{RF}$ calculated with the 2D RF power coupling model is decreasing. The increasing $L_\mathrm{net}$ dominates the decrease of $I_\mathrm{RF}$ in equation~(\ref{eq:U_winding}). However, since there is also the increasing number of coil windings in the denominator, the resulting $U_\mathrm{winding}$ decreases as well, as shown in figure~\ref{fig:NumberOfCoilWindings}(a).

The network resistance $R_\mathrm{net}$, as shown in figure~\ref{fig:NumberOfCoilWindings}(b), increases for two reasons. First, eddy currents in the Faraday shield increase as more of its area is covered by the RF coil and second, there is simply more Joule heating in a longer RF coil. The increase of $R_\mathrm{pl}$ (resulting from the decreased $I_\mathrm{RF}$) is stronger than the corresponding increase of $R_\mathrm{net}$, wherefore $\eta$ increases, as shown in figure~\ref{fig:NumberOfCoilWindings}(c). However, $\eta$ saturates for $N_\mathrm{windings} \geq 5$, because the decreasing $I_\mathrm{RF}^2$ and the increasing $R_\mathrm{net}$ almost fully compensate each other. For this reason the power absorbed in the RF network $P_\mathrm{net} = \frac{1}{2}R_\mathrm{net} I_\mathrm{RF}^2$ remains essentially constant.\cite{Briefi_2022} Hence for the current axial driver length a number of coil windings between five and eight is optimal, as already used in present setups and also foreseen for ITER.

\section{Optimizing the discharge geometry}
\label{sec:OptimizingDischargeGeometry}

\subsection{Driver axial length}
\label{subsec:DriverAxialLength}

\begin{figure}[h!]
	\centering
	\includegraphics[width=0.9\textwidth]{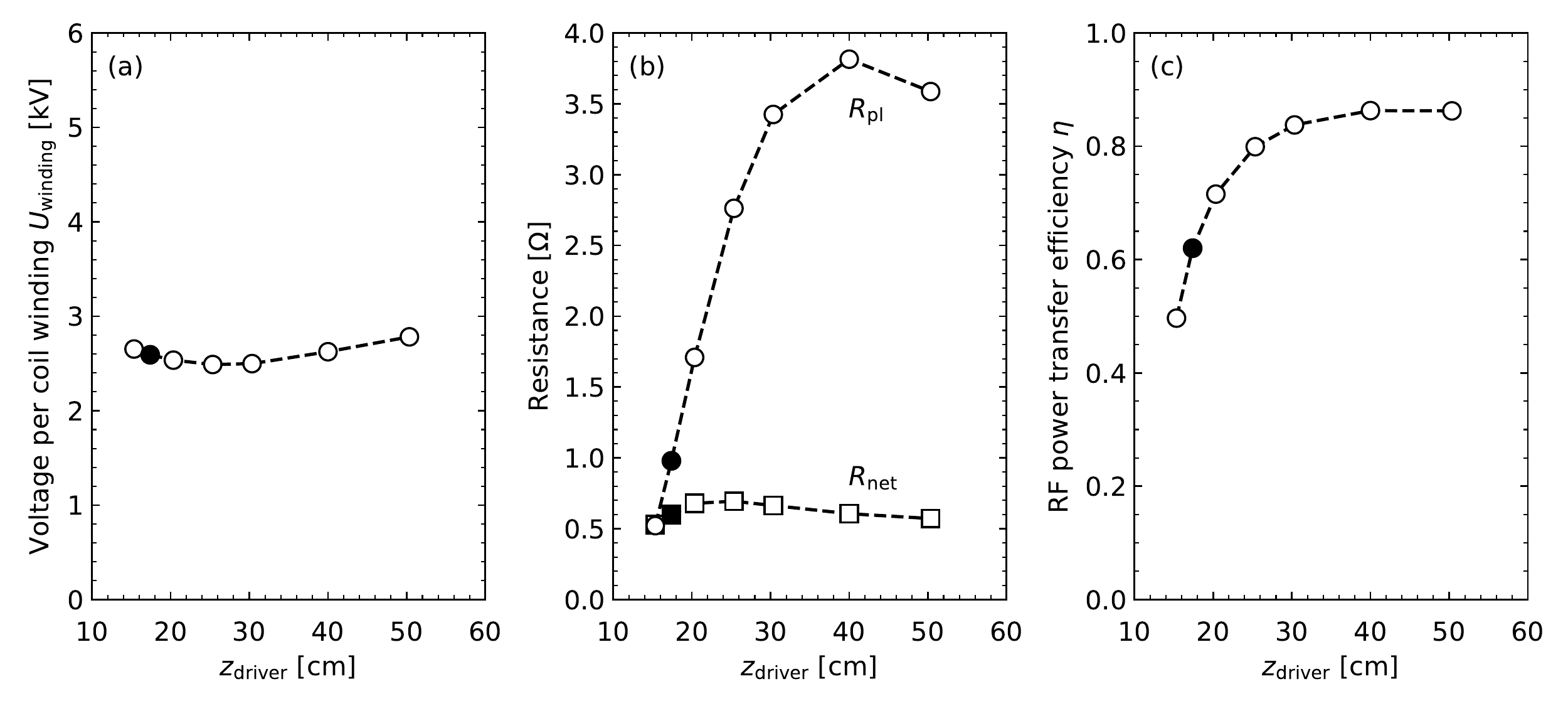}
	\caption{(a) Voltage per coil winding $U_\mathrm{winding}$, (b) equivalent plasma resistance $R_\mathrm{pl}$, network resistance $R_\mathrm{net}$ and (c) RF power transfer efficiency $\eta$ as a function of the axial driver length $z_\mathrm{driver}$. The full symbols indicate the validated values of the present experimental setup at $z_\mathrm{driver} = 17.4\,$cm. Figures (a) and (b) are adapted from Zielke,\cite{Zielke_thesis_2021} whereas (c) is reproduced from Briefi et al..\cite{Briefi_2022}}
	\label{fig:DriverAxialLength}
\end{figure}

\begin{figure}[h!]
	\centering
	\includegraphics[width=0.6\textwidth]{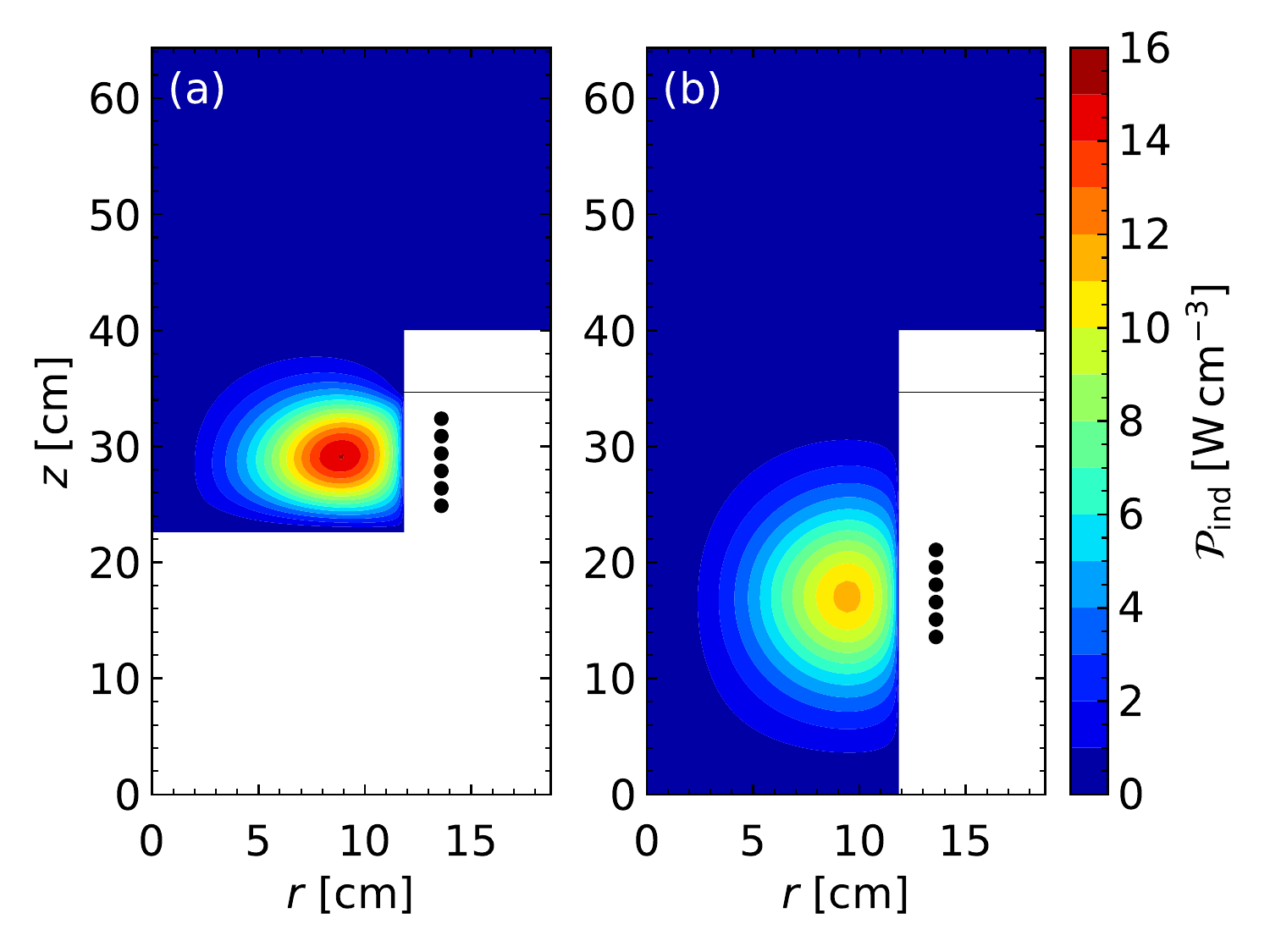}
	\caption{Calculated spatial profile of the inductive heating power $\mathcal{P}_\mathrm{ind}$ at a total delivered power of $P_\mathrm{RF}$ = 60\,kW and a filling pressure of 0.3\,Pa for the two different axial driver length (a) $z_\mathrm{driver} = 17.4\,$cm and (b) $z_\mathrm{driver} = 40\,$cm. Figure adapted from Zielke.\cite{Zielke_thesis_2021}}
	\label{fig:DriverAxialLength_2Dplots}
\end{figure}

Figure~\ref{fig:DriverAxialLength} shows the values of the modeled $U_\mathrm{winding}$, resistances and $\eta$ for a $z_\mathrm{driver}$, which is varied between 15 and 50\,cm. 
As shown in figure~\ref{fig:DriverAxialLength}(a), the voltage per coil winding is only slightly affected when the driver axial length is increased. The changing quantities in equation~(\ref{eq:U_winding}) are the network inductance $L_\mathrm{net}$ and the RF coil current $I_\mathrm{RF}$. The former increases from around 8\,$\mu$H at $z_\mathrm{driver} = 15\,$cm to 16\,$\mu$H at 50\,cm, as calculated by the source impedance model. Simultaneously the heated plasma volume becomes larger. This is apparent in figure~\ref{fig:DriverAxialLength_2Dplots}, where the power deposition profile of a short and a long driver are compared.
As can be seen, the absolute value of the absorbed power per unit volume is smaller in the case of the longer driver (11 vs.\ 15\,W\,cm$^{-3}$). However, the heated volume is increased considerably, wherefore more power is absorbed in total. Consequently, a lower RF coil current is needed at a larger axial driver length. The increase in $L_\mathrm{net}$ and decrease in $I_\mathrm{RF}$, which is also roughly a factor of two, effectively compensate each other, such that $U_\mathrm{winding}$ remains almost constant.

The larger power absorption at a lower RF current tends to increase $R_\mathrm{pl}$. The network resistance as shown in figure~\ref{fig:DriverAxialLength}(b) remains approximately constant. This has two reasons. First, the RF coil is not changed, wherefore the same amount of power is absorbed in it when the Faraday shield becomes longer. And second, while it is true that the eddy current distribution and the distribution of the absorbed power change with the length of the Faraday shield, the source impedance model shows that the total absorbed power remains approximately constant.

From the increasing $R_\mathrm{pl}$ and constant $R_\mathrm{net}$ follows a better RF power transfer efficiency $\eta$ according to equation~(\ref{eq:eta}). As shown in figure~\ref{fig:DriverAxialLength}(c), following the trend of $R_\mathrm{pl}$, $\eta$ saturates for $z_\mathrm{driver} > 30\,$cm, where $\eta\approx 85$\%. Hence $z_\mathrm{driver} = 30\,$cm is considered an optimal axial driver length. Also because of assembly space constraints and mechanical stability considerations it is advisable to not increase $z_\mathrm{driver}$ beyond 30\,cm.

To assess whether there are any drawbacks when $z_\mathrm{driver} = 30\,$cm is used, the electron density in the driver as well as the fluxes of neutral atoms and positive ions onto the plasma grid are calculated with the model. The results show that the volume averaged electron density does only decrease by around 1\%, when compared to the baseline configuration. The atomic flux onto the plasma grid is found to be only slightly reduced by 3\%, whereas the flux of positive ions, i.e.\ the sum of H$^+$, H$_2^+$ and H$_3^+$ is even increased by around 7\%. Hence no significant changes for the surface production of negative ions are expected. It has been checked that also at smaller and larger radii the trends as shown in figure~\ref{fig:DriverAxialLength} are persistent, i.e.~also at a different driver radius, it is beneficial to increase the axial driver length.

\subsection{Driver radius}
\label{subsec:DriverRadius}
\begin{figure}[h!]
	\centering
	\includegraphics[width=0.9\textwidth]{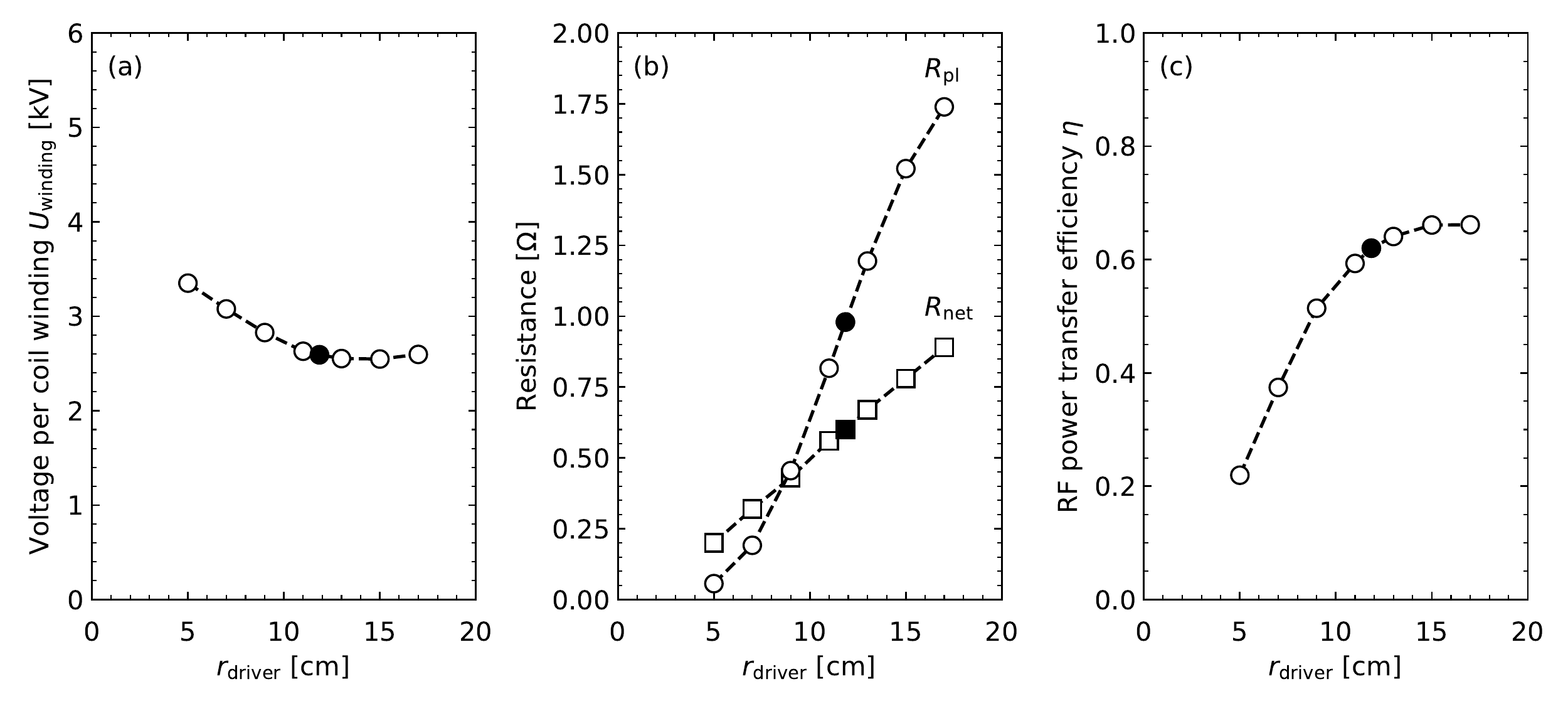}
	\caption{(a) Voltage per coil winding $U_\mathrm{winding}$, (b) equivalent plasma resistance $R_\mathrm{pl}$, network resistance $R_\mathrm{net}$ and (c) RF power transfer efficiency $\eta$ as a function of the driver radius $r_\mathrm{driver}$. The full symbols indicate the validated values of the present experimental setup at $r_\mathrm{driver} = 11.85\,$cm. Figures (a) and (b) are adapted from Zielke,\cite{Zielke_thesis_2021} whereas (c) is reproduced from Briefi et al..\cite{Briefi_2022}}
	\label{fig:DriverRadius}
\end{figure}

The larger magnetic flux produced by a coil with a bigger radius together with the increased Faraday shield surface lead to a slight increase of the network inductance $L_\mathrm{net}$. However, this is dominated by a decreasing $I_\mathrm{RF}$, as explained below. The resulting decrease of $U_\mathrm{winding}$ is shown in figure~\ref{fig:DriverRadius}(a).

\begin{figure}[h!]
	\centering
	\includegraphics[width=0.65\textwidth]{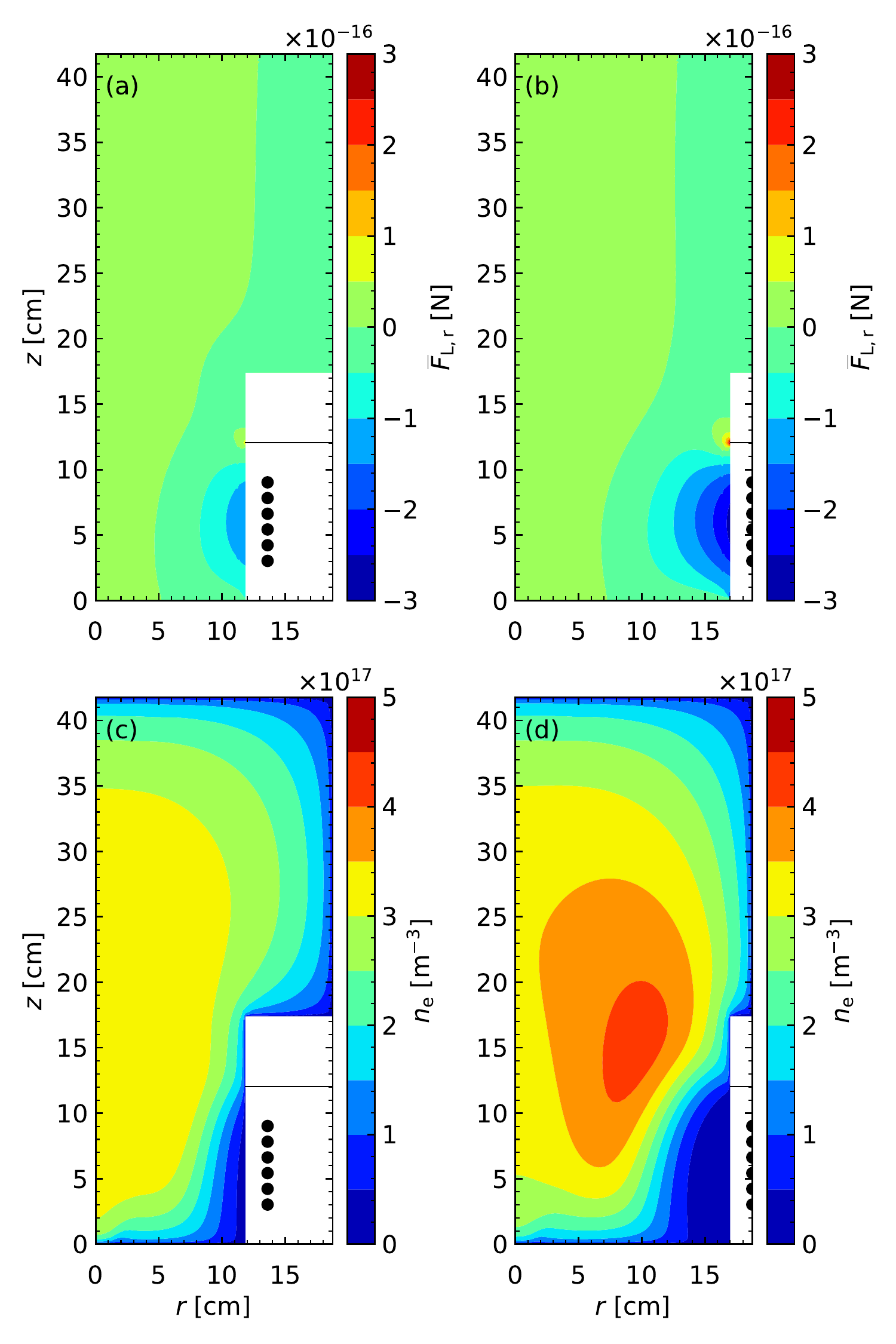}
	\caption{Calculated spatial profiles at $P_\mathrm{RF} = 60\,$kW and $p_\mathrm{fill} = 0.3\,$Pa for two different driver radii $r_\mathrm{driver} = 11.85\,$cm (see (a) and (c)) and $r_\mathrm{driver} = 17\,$cm (see (b) and (d)), respectively. Plots (a) and (b) show the spatial profiles of the radial component of the RF-averaged Lorentz force $\bar{F}_{\mathrm{L},r}$, whereas (c) and (d) show the corresponding electron densities $n_\mathrm{e}$. Figure adapted from Zielke.\cite{Zielke_thesis_2021}}
	\label{fig:DriverRadius_2Dplots}
\end{figure}

Because of the larger Faraday shield surface and longer RF coil the network resistance $R_\mathrm{net}$ in figure~\ref{fig:DriverRadius}(b) increases. The plasma equivalent resistance $R_\mathrm{pl}$ becomes also larger, simply because of the bigger plasma volume, where more electrons can be heated. Simultaneously the self-consistent $I_\mathrm{RF}$ decreases. However, this effect is not as pronounced as in the case of the axial driver length, because the RF-averaged Lorentz force (ponderomotive force) tends to push the plasma away from the RF coil. This effect is more pronounced at larger radii, as shown in figure~\ref{fig:DriverRadius_2Dplots}.

In the driver the radial component of the RF Lorentz force is mainly negative, thus pointing inwards, as shown in figure~\ref{fig:DriverRadius_2Dplots}(a) for a smaller radius and in (b) for a larger one, respectively. At an increased radius the magnitude as well as how far the RF Lorentz force reaches into the driver is larger by roughly a factor of two. This force affects the plasma density profile in the driver, which is more depleted when $r_\mathrm{driver}$ is larger, as becomes evident from a comparison of figure~\ref{fig:DriverRadius_2Dplots}(c) and (d). As a consequence, less electrons are heated. For this reason $\eta$ cannot be increased significantly beyond its current value of around 60\% at $r_\mathrm{driver} = 11.85\,$cm, as shown in figure~\ref{fig:DriverRadius}(c).

\section{Optimizing the driving frequency}
\label{sec:OptimizingRF}

For the optimization of the driving frequency the power absorbed by the plasma $P_\mathrm{pl}$ is fixed at $P_\mathrm{pl} = 30\,$kW ($p_\mathrm{fill} = 0.3\,$Pa as usual). Note that also at the experimental setup $P_\mathrm{RF}$ is adjusted, such that a certain $P_\mathrm{pl}$ (which determines the plasma parameters in the driver)\cite{Zielke_2021} is achieved. In this way the calculated values of $U_\mathrm{winding}$ for the various driving frequencies are better comparable to their experimental counterparts.

\begin{figure}[h!]
	\centering
	\includegraphics[width=0.9\textwidth]{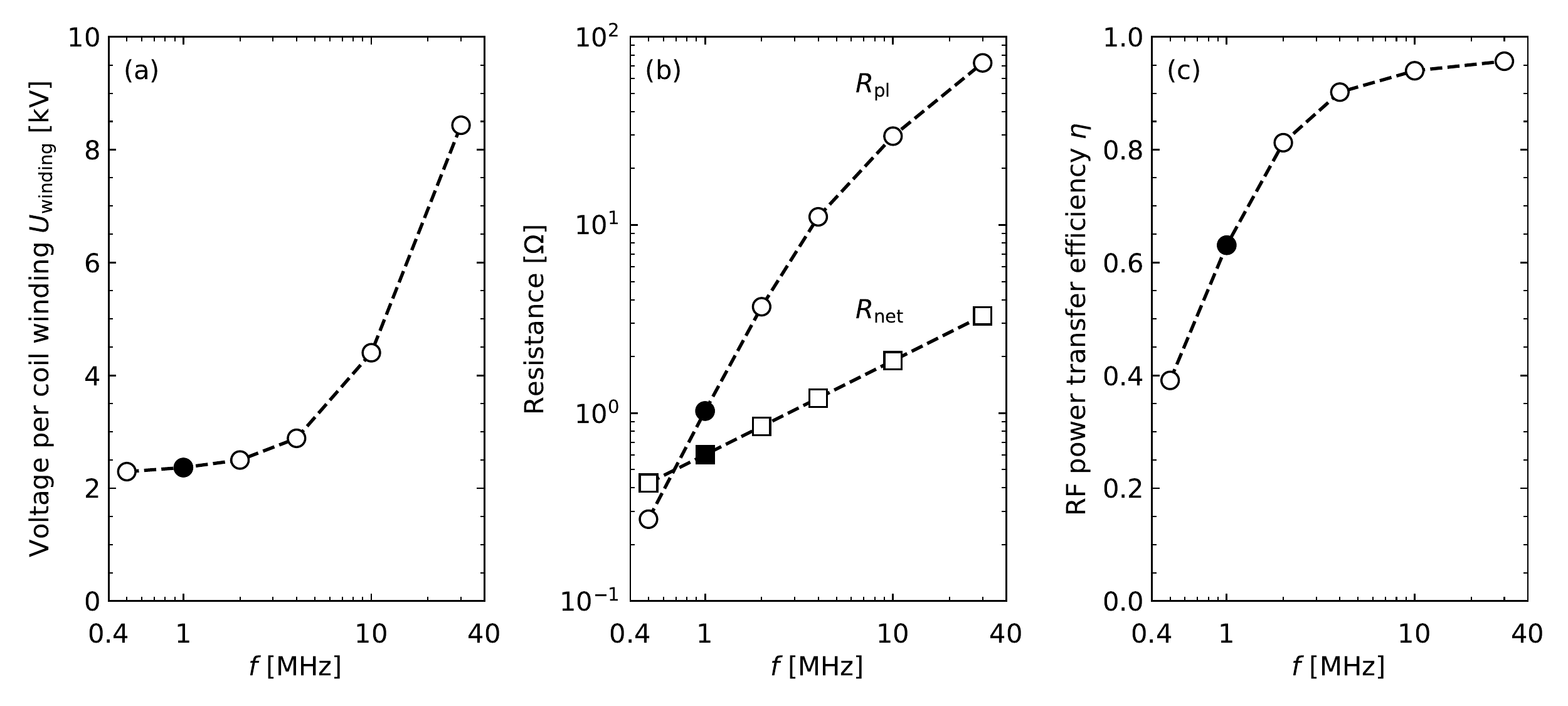}
	\caption{(a) Voltage per coil winding $U_\mathrm{winding}$, (b) equivalent plasma resistance $R_\mathrm{pl}$, network resistance $R_\mathrm{net}$ and (c) RF power transfer efficiency $\eta$ as a function of the driving frequency $f$. The full symbols indicate the validated values of the present experimental setup at $f = 1\,$MHz. Figure adapted from Zielke.\cite{Zielke_thesis_2021}}
	\label{fig:RF}
\end{figure}
The voltage per coil winding increases for a larger driving frequency (see figure~\ref{fig:RF}(a)) mainly because $U_\mathrm{winding} \propto f$, as stated in equation~(\ref{eq:U_winding}). To sustain the same electric RF field $\bm{E}$ in the plasma Faraday's law states that at higher applied frequency $\omega$ a lower magnetic field $\bm{B}$ is necessary, since $E \propto \omega B$. According to Amp\`{e}re's law this entails lower RF coil currents, wherefore the plasma resistance is increased at a constant $P_\mathrm{pl} = \frac{1}{2} R_\mathrm{pl} I_\mathrm{RF}^2$. As can be seen in figure~\ref{fig:RF}(b), the increase in $R_\mathrm{pl}$ is almost two orders of magnitude, when e.g.\ 1\,MHz is compared to 30\,MHz. Because of the more pronounced skin effect in the RF coil and Faraday shield at higher driving frequency, the network resistance $R_\mathrm{net} \propto \omega^{1/2}$ increases as well. Quantitatively however, the increase in $R_\mathrm{pl}$ dominates the one in $R_\mathrm{net}$ by far, wherefore also $\eta$ increases considerably, as shown in figure~\ref{fig:RF}(c). In other words: to make the plasma absorb 30\,kW only 32\,kW of generator power is needed at $f > 10\,$MHz, whereas at $f = 0.5\,$MHz 75\,kW of generator power is necessary. However, there is also the drawback of an increasing $U_\mathrm{winding}$, when the driving frequency is increased. This becomes considerable for $f > 2\,$MHz, wherefore a conservative optimum is at 2\,MHz.\cite{Briefi_2022} Here, $\eta$ is increased from 60\% to slightly above 80\%, whereas $U_\mathrm{winding}$ remains almost constant at around 2.5\,kV.

At $f = 2$\,MHz the volume averaged electron density in the driver as well as the fluxes of atoms and positive ions onto the plasma grid are essentially unchanged when compared to the baseline case at $f = 1\,$MHz. Hence no drawbacks regarding the surface production of negative ions are expected.

\section{Combining longer driver and increased driving frequency}
\label{sec:GlobalOptimum}

For the ITER prototype RF ion source two promising measures for optimizing the RF power coupling have been identified in sections~\ref{subsec:DriverAxialLength} and~\ref{sec:OptimizingRF}. These are a longer axial driver with 30 instead of 18\,cm and a higher driving frequency of 2 instead of 1\,MHz. In the following study these two measures are combined. The results are summarized in table~\ref{tab:GlobalOptimization}.
\begin{table}[h!]
	\caption{Further optimization of the RF power coupling at the ITER prototype RF ion source at the Batman Upgrade testbed. The fixed power absorbed by the plasma is $P_\mathrm{pl} = 30\,$kW and the ITER relevant filling pressure $p_\mathrm{fill} = 0.3\,$Pa.}
	\label{tab:GlobalOptimization}
	\centering
	\small{
		\begin{ruledtabular}
		\begin{tabular}{ccccc}
			& baseline & $z_\mathrm{driver}$ optimized & $f$ optimized & $z_\mathrm{driver}$ \& $f$ optimized\\ 
			\hline
			$U_\mathrm{winding}$ [kV] & 2.4 & 2.2 & 2.5 & 2.3 \\
			$\eta$ & 0.63 & 0.81 & 0.81 & 0.91 \\
		\end{tabular}
		\end{ruledtabular}
	}
\end{table}

The longer driver leads to an increased plasma volume, whereas the RF coil current and the ponderomotive effect are decreased at the larger frequency. The interplay of these effects yields a nonlinear increase of the plasma resistance and $\eta$, when the two measures are combined. In this way it is possible to considerably increase $\eta$ from 63\% to 91\%, while holding $U_\mathrm{winding}$ essentially constant. Note that for the above comparison $P_\mathrm{pl} = 30\,$kW was fixed. Therefore, in the case when $z_\mathrm{driver}$ and $f$ are optimized the average electron density decreases by -10\% and the fluxes of H$^+$, H$_2^+$, H$_3^+$ and H onto the plasma grid decrease by around -28\%, -13\%, -8\% and -12\%, respectively, when compared to the baseline case.

However, when the optimization studies are performed at a fixed generator power of 60\,kW, then the electron density averaged in the driver increases by around +8\%, and the respective fluxes change by +11\%, +11\%, -18\% and -1\%. In the baseline case, only around 38\,kW are absorbed by the plasma, whereas in the case when $z_\mathrm{driver}$ and $f$ are optimized, around 55\,kW are absorbed. In the latter case the coil voltage per winding increases to 3.3\,kV compared to 2.7\,kV in the baseline case. Nevertheless it is expected that the chance for an RF breakdown does not increase considerably, especially since in the optimized case lower generator powers can be used to obtain the same plasma power, as shown above. The accompanying overall increase in driver electron density as well as in the dominant fluxes onto the plasma grid (which are the ones of H$^+$ and H$_2^+$ at 0.3\,Pa) shows, that in the optimized case a positive effect on the surface production of negative ions can be expected.

\section{Conclusion}
\label{sec:Conclusion}

A self-consistent modeling approach, which combines a 3D source impedance model and a 2D cylindrically symmetric fluid-electromagnetic model, is used for simulation and optimization of the RF power coupling in ITER-relevant negative hydrogen RF ion sources. In the RF network, the losses are found to be determined by changing conductor surfaces, where eddy currents are driven, and by the skin effect, which changes with the applied frequency. On the plasma side there are linear effects, such as a lower RF coil current needed at higher driving frequency to generate the electric RF field in the plasma. And there are nonlinear plasma characteristics, such as the RF Lorentz force, which is strongly affected by the driver geometry. At a fixed radius a larger axial driver length is shown to be highly beneficial, since it increases the plasma volume, but leaves the RF Lorentz force unchanged. Therefore more electrons can be heated. However, when the driver radius is increased at fixed axial length, the beneficial effect of a bigger volume is diminished by a nonlinearly increasing RF Lorentz force, leading to a depletion of electrons in the driver.

For the source impedance model used to calculate the network impedance no further improvements regarding the implemented physics are necessary. However, extending the formulation of the RF power coupling model (which is currently 2D with the assumption of cylindrical symmetry) towards 3D would allow to fully implement the 3D topologies of the magnetic fields. In particular the one of the magnetic filter field, such that particle transport in the expansion chamber can be studied. Also 3D driver geometries, such as the racetrack-shaped driver,\cite{Fantz_2018} as presently suggested for the European pulsed DEMOnstration Power Plant (DEMO),\cite{Tran_2022} would become possible. In a first approach the 2D cylindrically symmetric model showed, that the scaling with the axial driver length still holds at an increased driver radius (which is a reasonable first approximation for a racetrack-shaped driver). Subsequently, a 3D implementation of the self-consistent fluid-electromagnetic model is currently in preparation.

The obtained results predict that for the ITER prototype RF negative ion source the RF power transfer efficiency can be significantly improved from 60\% to 90\% by increasing the axial driver length and driving frequency. In this way the network losses are drastically reduced making possible lower required generator powers as well as a decreased effort for the cooling of the RF coil and the Faraday shield. At the same time, the voltage at the RF coil remains essentially constant, wherefore the probability for electrical arcs does not increase by the proposed measures. Consequently the RF ion source operates more efficiently and reliably. For all proposed measures beneficial for the RF power coupling, the model calculates that indicators for the performance of the negative ion source such as plasma density in the driver and the fluxes of atoms and positive ions onto the plasma grid are not negatively affected.

\begin{acknowledgments}
\noindent{This work has been carried out within the framework of the EUROfusion Consortium and has received funding from the Euratom research and training programme 2014-2018 and 2019-2020 under grant agreement No 633053. The views and opinions expressed herein do not necessarily reflect those of the European Commission.}

\noindent{This work has been carried out within the framework of the EUROfusion Consortium, funded by the European Union via the Euratom Research and Training Programme (Grant Agreement No 101052200 — EUROfusion). Views and opinions expressed are however those of the author(s) only and do not necessarily reflect those of the European Union or the European Commission. Neither the European Union nor the European Commission can be held responsible for them.}
\end{acknowledgments}
		
\section*{Data Availability Statement}
		
\noindent{The data that support the findings of this study are available from the corresponding author upon reasonable request.}

\nocite{*}
\bibliography{library}

\begin{thebibliography}{19}%
\makeatletter
\providecommand \@ifxundefined [1]{%
 \@ifx{#1\undefined}
}%
\providecommand \@ifnum [1]{%
 \ifnum #1\expandafter \@firstoftwo
 \else \expandafter \@secondoftwo
 \fi
}%
\providecommand \@ifx [1]{%
 \ifx #1\expandafter \@firstoftwo
 \else \expandafter \@secondoftwo
 \fi
}%
\providecommand \natexlab [1]{#1}%
\providecommand \enquote  [1]{``#1''}%
\providecommand \bibnamefont  [1]{#1}%
\providecommand \bibfnamefont [1]{#1}%
\providecommand \citenamefont [1]{#1}%
\providecommand \href@noop [0]{\@secondoftwo}%
\providecommand \href [0]{\begingroup \@sanitize@url \@href}%
\providecommand \@href[1]{\@@startlink{#1}\@@href}%
\providecommand \@@href[1]{\endgroup#1\@@endlink}%
\providecommand \@sanitize@url [0]{\catcode `\\12\catcode `\$12\catcode
  `\&12\catcode `\#12\catcode `\^12\catcode `\_12\catcode `\%12\relax}%
\providecommand \@@startlink[1]{}%
\providecommand \@@endlink[0]{}%
\providecommand \url  [0]{\begingroup\@sanitize@url \@url }%
\providecommand \@url [1]{\endgroup\@href {#1}{\urlprefix }}%
\providecommand \urlprefix  [0]{URL }%
\providecommand \Eprint [0]{\href }%
\providecommand \doibase [0]{http://dx.doi.org/}%
\providecommand \selectlanguage [0]{\@gobble}%
\providecommand \bibinfo  [0]{\@secondoftwo}%
\providecommand \bibfield  [0]{\@secondoftwo}%
\providecommand \translation [1]{[#1]}%
\providecommand \BibitemOpen [0]{}%
\providecommand \bibitemStop [0]{}%
\providecommand \bibitemNoStop [0]{.\EOS\space}%
\providecommand \EOS [0]{\spacefactor3000\relax}%
\providecommand \BibitemShut  [1]{\csname bibitem#1\endcsname}%
\let\auto@bib@innerbib\@empty
\bibitem [{ITE()}]{ITER}%
  \BibitemOpen
  \href@noop {} {}\bibinfo {howpublished} {\url{www.iter.org}}\BibitemShut
  {NoStop}%
\bibitem [{\citenamefont {Hemsworth}\ \emph {et~al.}(2017)\citenamefont
  {Hemsworth}, \citenamefont {Boilson}, \citenamefont {Blatchford},
  \citenamefont {Palma}, \citenamefont {Chitarin}, \citenamefont {de~Esch},
  \citenamefont {Geli}, \citenamefont {Dremel}, \citenamefont {Graceffa},
  \citenamefont {Marcuzzi} \emph {et~al.}}]{Hemsworth_2017}%
  \BibitemOpen
  \bibfield  {author} {\bibinfo {author} {\bibfnamefont {R.~S.}\ \bibnamefont
  {Hemsworth}}, \bibinfo {author} {\bibfnamefont {D.}~\bibnamefont {Boilson}},
  \bibinfo {author} {\bibfnamefont {P.}~\bibnamefont {Blatchford}}, \bibinfo
  {author} {\bibfnamefont {M.~D.}\ \bibnamefont {Palma}}, \bibinfo {author}
  {\bibfnamefont {G.}~\bibnamefont {Chitarin}}, \bibinfo {author}
  {\bibfnamefont {H.~P.~L.}\ \bibnamefont {de~Esch}}, \bibinfo {author}
  {\bibfnamefont {F.}~\bibnamefont {Geli}}, \bibinfo {author} {\bibfnamefont
  {M.}~\bibnamefont {Dremel}}, \bibinfo {author} {\bibfnamefont
  {J.}~\bibnamefont {Graceffa}}, \bibinfo {author} {\bibfnamefont
  {D.}~\bibnamefont {Marcuzzi}},  \emph {et~al.},\ }\bibfield  {title}
  {\enquote {\bibinfo {title} {Overview of the design of the {ITER} heating
  neutral beam injectors},}\ }\href {\doibase 10.1088/1367-2630/19/2/025005}
  {\bibfield  {journal} {\bibinfo  {journal} {New Journal of Physics}\ }\textbf
  {\bibinfo {volume} {19}},\ \bibinfo {pages} {025005} (\bibinfo {year}
  {2017})}\BibitemShut {NoStop}%
\bibitem [{\citenamefont {Heinemann}\ \emph {et~al.}(2017)\citenamefont
  {Heinemann}, \citenamefont {Fantz}, \citenamefont {Kraus}, \citenamefont
  {Schiesko}, \citenamefont {Wimmer}, \citenamefont {Wünderlich},
  \citenamefont {Bonomo}, \citenamefont {Fröschle}, \citenamefont
  {Nocentini},\ and\ \citenamefont {Riedl}}]{Heinemann_2017}%
  \BibitemOpen
  \bibfield  {author} {\bibinfo {author} {\bibfnamefont {B.}~\bibnamefont
  {Heinemann}}, \bibinfo {author} {\bibfnamefont {U.}~\bibnamefont {Fantz}},
  \bibinfo {author} {\bibfnamefont {W.}~\bibnamefont {Kraus}}, \bibinfo
  {author} {\bibfnamefont {L.}~\bibnamefont {Schiesko}}, \bibinfo {author}
  {\bibfnamefont {C.}~\bibnamefont {Wimmer}}, \bibinfo {author} {\bibfnamefont
  {D.}~\bibnamefont {Wünderlich}}, \bibinfo {author} {\bibfnamefont
  {F.}~\bibnamefont {Bonomo}}, \bibinfo {author} {\bibfnamefont
  {M.}~\bibnamefont {Fröschle}}, \bibinfo {author} {\bibfnamefont
  {R.}~\bibnamefont {Nocentini}}, \ and\ \bibinfo {author} {\bibfnamefont
  {R.}~\bibnamefont {Riedl}},\ }\bibfield  {title} {\enquote {\bibinfo {title}
  {Towards large and powerful radio frequency driven negative ion sources for
  fusion},}\ }\href {\doibase 10.1088/1367-2630/aa520c} {\bibfield  {journal}
  {\bibinfo  {journal} {New Journal of Physics}\ }\textbf {\bibinfo {volume}
  {19}},\ \bibinfo {pages} {015001} (\bibinfo {year} {2017})}\BibitemShut
  {NoStop}%
\bibitem [{\citenamefont {Zielke}\ \emph {et~al.}(2022)\citenamefont {Zielke},
  \citenamefont {Briefi}, \citenamefont {Lishev},\ and\ \citenamefont
  {Fantz}}]{Zielke_2022}%
  \BibitemOpen
  \bibfield  {author} {\bibinfo {author} {\bibfnamefont {D.}~\bibnamefont
  {Zielke}}, \bibinfo {author} {\bibfnamefont {S.}~\bibnamefont {Briefi}},
  \bibinfo {author} {\bibfnamefont {S.}~\bibnamefont {Lishev}}, \ and\ \bibinfo
  {author} {\bibfnamefont {U.}~\bibnamefont {Fantz}},\ }\bibfield  {title}
  {\enquote {\bibinfo {title} {Modeling inductive radio frequency coupling in
  powerful negative hydrogen ion sources: validating a self-consistent ﬂuid
  model},}\ }\href {\doibase 10.1088/1361-6595/ac5845} {\bibfield  {journal}
  {\bibinfo  {journal} {Plasma Sources Science and Technology}\ }\textbf
  {\bibinfo {volume} {31}},\ \bibinfo {pages} {035019} (\bibinfo {year}
  {2022})}\BibitemShut {NoStop}%
\bibitem [{\citenamefont {Krylov}\ and\ \citenamefont
  {Hemsworth}(2006)}]{Krylov_2006}%
  \BibitemOpen
  \bibfield  {author} {\bibinfo {author} {\bibfnamefont {A.}~\bibnamefont
  {Krylov}}\ and\ \bibinfo {author} {\bibfnamefont {R.}~\bibnamefont
  {Hemsworth}},\ }\bibfield  {title} {\enquote {\bibinfo {title} {{Gas flow and
  related beam losses in the ITER neutral beam injector}},}\ }\href {\doibase
  https://doi.org/10.1016/j.fusengdes.2006.03.006} {\bibfield  {journal}
  {\bibinfo  {journal} {Fusion Engineering and Design}\ }\textbf {\bibinfo
  {volume} {81}},\ \bibinfo {pages} {2239--2248} (\bibinfo {year}
  {2006})}\BibitemShut {NoStop}%
\bibitem [{\citenamefont {Fantz}\ \emph {et~al.}(2017)\citenamefont {Fantz},
  \citenamefont {Hopf}, \citenamefont {Wünderlich}, \citenamefont {Friedl},
  \citenamefont {Fröschle}, \citenamefont {Heinemann}, \citenamefont {Kraus},
  \citenamefont {Kurutz}, \citenamefont {Riedl}, \citenamefont {Nocentini},\
  and\ \citenamefont {Schiesko}}]{Fantz_2017}%
  \BibitemOpen
  \bibfield  {author} {\bibinfo {author} {\bibfnamefont {U.}~\bibnamefont
  {Fantz}}, \bibinfo {author} {\bibfnamefont {C.}~\bibnamefont {Hopf}},
  \bibinfo {author} {\bibfnamefont {D.}~\bibnamefont {Wünderlich}}, \bibinfo
  {author} {\bibfnamefont {R.}~\bibnamefont {Friedl}}, \bibinfo {author}
  {\bibfnamefont {M.}~\bibnamefont {Fröschle}}, \bibinfo {author}
  {\bibfnamefont {B.}~\bibnamefont {Heinemann}}, \bibinfo {author}
  {\bibfnamefont {W.}~\bibnamefont {Kraus}}, \bibinfo {author} {\bibfnamefont
  {U.}~\bibnamefont {Kurutz}}, \bibinfo {author} {\bibfnamefont
  {R.}~\bibnamefont {Riedl}}, \bibinfo {author} {\bibfnamefont
  {R.}~\bibnamefont {Nocentini}}, \ and\ \bibinfo {author} {\bibfnamefont
  {L.}~\bibnamefont {Schiesko}},\ }\bibfield  {title} {\enquote {\bibinfo
  {title} {Towards powerful negative ion beams at the test facility {ELISE} for
  the {ITER} and {DEMO} {NBI} systems},}\ }\href {\doibase
  10.1088/1741-4326/aa778b} {\bibfield  {journal} {\bibinfo  {journal} {Nuclear
  Fusion}\ }\textbf {\bibinfo {volume} {57}},\ \bibinfo {pages} {116007}
  (\bibinfo {year} {2017})}\BibitemShut {NoStop}%
\bibitem [{\citenamefont {Zielke}, \citenamefont {Briefi},\ and\ \citenamefont
  {Fantz}(2021)}]{Zielke_2021}%
  \BibitemOpen
  \bibfield  {author} {\bibinfo {author} {\bibfnamefont {D.}~\bibnamefont
  {Zielke}}, \bibinfo {author} {\bibfnamefont {S.}~\bibnamefont {Briefi}}, \
  and\ \bibinfo {author} {\bibfnamefont {U.}~\bibnamefont {Fantz}},\ }\bibfield
   {title} {\enquote {\bibinfo {title} {{RF} power transfer efficiency and
  plasma parameters of low pressure high power {ICPs}},}\ }\href {\doibase
  10.1088/1361-6463/abd8ee} {\bibfield  {journal} {\bibinfo  {journal} {Journal
  of Physics D: Applied Physics}\ }\textbf {\bibinfo {volume} {54}},\ \bibinfo
  {pages} {155202} (\bibinfo {year} {2021})}\BibitemShut {NoStop}%
\bibitem [{\citenamefont {Toigo}\ \emph {et~al.}(2021)\citenamefont {Toigo},
  \citenamefont {Marcuzzi}, \citenamefont {Serianni}, \citenamefont {Boldrin},
  \citenamefont {Chitarin}, \citenamefont {Bello}, \citenamefont {Grando},
  \citenamefont {Luchetta}, \citenamefont {Pasqualotto}, \citenamefont
  {Zaccaria} \emph {et~al.}}]{Toigo_2021}%
  \BibitemOpen
  \bibfield  {author} {\bibinfo {author} {\bibfnamefont {V.}~\bibnamefont
  {Toigo}}, \bibinfo {author} {\bibfnamefont {D.}~\bibnamefont {Marcuzzi}},
  \bibinfo {author} {\bibfnamefont {G.}~\bibnamefont {Serianni}}, \bibinfo
  {author} {\bibfnamefont {M.}~\bibnamefont {Boldrin}}, \bibinfo {author}
  {\bibfnamefont {G.}~\bibnamefont {Chitarin}}, \bibinfo {author}
  {\bibfnamefont {S.~D.}\ \bibnamefont {Bello}}, \bibinfo {author}
  {\bibfnamefont {L.}~\bibnamefont {Grando}}, \bibinfo {author} {\bibfnamefont
  {A.}~\bibnamefont {Luchetta}}, \bibinfo {author} {\bibfnamefont
  {R.}~\bibnamefont {Pasqualotto}}, \bibinfo {author} {\bibfnamefont
  {P.}~\bibnamefont {Zaccaria}},  \emph {et~al.},\ }\bibfield  {title}
  {\enquote {\bibinfo {title} {{On the road to ITER NBIs: SPIDER improvement
  after first operation and MITICA construction progress}},}\ }\href {\doibase
  https://doi.org/10.1016/j.fusengdes.2021.112622} {\bibfield  {journal}
  {\bibinfo  {journal} {Fusion Engineering and Design}\ }\textbf {\bibinfo
  {volume} {168}},\ \bibinfo {pages} {112622} (\bibinfo {year}
  {2021})}\BibitemShut {NoStop}%
\bibitem [{\citenamefont {Jain}\ \emph {et~al.}(2022)\citenamefont {Jain},
  \citenamefont {Recchia}, \citenamefont {Maistrello},\ and\ \citenamefont
  {Gaio}}]{Jain_2022}%
  \BibitemOpen
  \bibfield  {author} {\bibinfo {author} {\bibfnamefont {P.}~\bibnamefont
  {Jain}}, \bibinfo {author} {\bibfnamefont {M.}~\bibnamefont {Recchia}},
  \bibinfo {author} {\bibfnamefont {A.}~\bibnamefont {Maistrello}}, \ and\
  \bibinfo {author} {\bibfnamefont {E.}~\bibnamefont {Gaio}},\ }\bibfield
  {title} {\enquote {\bibinfo {title} {{Investigation of RF driver equivalent
  impedance in the inductively coupled SPIDER ion source}},}\ }\href
  {http://iopscience.iop.org/article/10.1088/1361-6587/ac8617} {\bibfield
  {journal} {\bibinfo  {journal} {Plasma Physics and Controlled Fusion}\ }
  (\bibinfo {year} {2022})}\BibitemShut {NoStop}%
\bibitem [{\citenamefont {Hagelaar}, \citenamefont {Fubiani},\ and\
  \citenamefont {Boeuf}(2011)}]{Hagelaar_2011}%
  \BibitemOpen
  \bibfield  {author} {\bibinfo {author} {\bibfnamefont {G.~J.~M.}\
  \bibnamefont {Hagelaar}}, \bibinfo {author} {\bibfnamefont {G.}~\bibnamefont
  {Fubiani}}, \ and\ \bibinfo {author} {\bibfnamefont {J.~P.}\ \bibnamefont
  {Boeuf}},\ }\bibfield  {title} {\enquote {\bibinfo {title} {Model of an
  inductively coupled negative ion source: I. general model description},}\
  }\href {\doibase 10.1088/0963-0252/20/1/015001} {\bibfield  {journal}
  {\bibinfo  {journal} {Plasma Sources Sci. Technol.}\ }\textbf {\bibinfo
  {volume} {20}},\ \bibinfo {pages} {015001} (\bibinfo {year}
  {2011})}\BibitemShut {NoStop}%
\bibitem [{\citenamefont {Jain}\ \emph {et~al.}(2018)\citenamefont {Jain},
  \citenamefont {Recchia}, \citenamefont {Cavenago}, \citenamefont {Fantz},
  \citenamefont {Gaio}, \citenamefont {Kraus}, \citenamefont {Maistrello},\
  and\ \citenamefont {Veltri}}]{Jain_2018_1}%
  \BibitemOpen
  \bibfield  {author} {\bibinfo {author} {\bibfnamefont {P.}~\bibnamefont
  {Jain}}, \bibinfo {author} {\bibfnamefont {M.}~\bibnamefont {Recchia}},
  \bibinfo {author} {\bibfnamefont {M.}~\bibnamefont {Cavenago}}, \bibinfo
  {author} {\bibfnamefont {U.}~\bibnamefont {Fantz}}, \bibinfo {author}
  {\bibfnamefont {E.}~\bibnamefont {Gaio}}, \bibinfo {author} {\bibfnamefont
  {W.}~\bibnamefont {Kraus}}, \bibinfo {author} {\bibfnamefont
  {A.}~\bibnamefont {Maistrello}}, \ and\ \bibinfo {author} {\bibfnamefont
  {P.}~\bibnamefont {Veltri}},\ }\bibfield  {title} {\enquote {\bibinfo {title}
  {Evaluation of power transfer efficiency for a high power inductively coupled
  radio-frequency hydrogen ion source},}\ }\href {\doibase
  10.1088/1361-6587/aaab19} {\bibfield  {journal} {\bibinfo  {journal} {Plasma
  Physics and Controlled Fusion}\ }\textbf {\bibinfo {volume} {60}},\ \bibinfo
  {pages} {045007} (\bibinfo {year} {2018})}\BibitemShut {NoStop}%
\bibitem [{\citenamefont {Chen}\ \emph {et~al.}(2021)\citenamefont {Chen},
  \citenamefont {Li}, \citenamefont {Zuo}, \citenamefont {Li},\ and\
  \citenamefont {Chen}}]{Chen_2021}%
  \BibitemOpen
  \bibfield  {author} {\bibinfo {author} {\bibfnamefont {P.}~\bibnamefont
  {Chen}}, \bibinfo {author} {\bibfnamefont {D.}~\bibnamefont {Li}}, \bibinfo
  {author} {\bibfnamefont {C.}~\bibnamefont {Zuo}}, \bibinfo {author}
  {\bibfnamefont {Z.}~\bibnamefont {Li}}, \ and\ \bibinfo {author}
  {\bibfnamefont {D.}~\bibnamefont {Chen}},\ }\bibfield  {title} {\enquote
  {\bibinfo {title} {{A method to evaluate the plasma equivalent resistance of
  fusion relevant RF ion sources}},}\ }\href {\doibase
  https://doi.org/10.1016/j.fusengdes.2021.112926} {\bibfield  {journal}
  {\bibinfo  {journal} {Fusion Engineering and Design}\ }\textbf {\bibinfo
  {volume} {173}},\ \bibinfo {pages} {112926} (\bibinfo {year}
  {2021})}\BibitemShut {NoStop}%
\bibitem [{\citenamefont {Zielke}(2021)}]{Zielke_thesis_2021}%
  \BibitemOpen
  \bibfield  {author} {\bibinfo {author} {\bibfnamefont {D.}~\bibnamefont
  {Zielke}},\ }\emph {\bibinfo {title} {Development of a predictive
  self-consistent fluid model for optimizing inductive RF coupling of powerful
  negative hydrogen ion sources}},\ \href
  {https://nbn-resolving.org/urn:nbn:de:bvb:384-opus4-894170} {Ph.D. thesis},\
  \bibinfo  {school} {Augsburg University} (\bibinfo {year} {2021})\BibitemShut
  {NoStop}%
\bibitem [{\citenamefont {Briefi}, \citenamefont {Zielke},\ and\ \citenamefont
  {Fantz}(2023)}]{Briefi_2023}%
  \BibitemOpen
  \bibfield  {author} {\bibinfo {author} {\bibfnamefont {S.}~\bibnamefont
  {Briefi}}, \bibinfo {author} {\bibfnamefont {D.}~\bibnamefont {Zielke}}, \
  and\ \bibinfo {author} {\bibfnamefont {U.}~\bibnamefont {Fantz}},\ }\bibfield
   {title} {\enquote {\bibinfo {title} {{Determining RF network losses at the
  H$^-$ ion source test bed BATMAN Upgrade via 3D electromagnetic modeling}},}\
  }\href@noop {} {\bibfield  {journal} {\bibinfo  {journal} {in preparation}\ }
  (\bibinfo {year} {2023})}\BibitemShut {NoStop}%
\bibitem [{\citenamefont {{COMSOL AB, Stockholm, Sweden}}(2021)}]{comsol_60}%
  \BibitemOpen
  \bibfield  {author} {\bibinfo {author} {\bibnamefont {{COMSOL AB, Stockholm,
  Sweden}}},\ }\href@noop {} {\enquote {\bibinfo {title} {Comsol
  multiphysics\textsuperscript{\textregistered}, version 6.0},}\ } (\bibinfo
  {year} {2021})\BibitemShut {NoStop}%
\bibitem [{\citenamefont {Froese}, \citenamefont {Smolyakov},\ and\
  \citenamefont {Sydorenko}(2009)}]{Froese_2009}%
  \BibitemOpen
  \bibfield  {author} {\bibinfo {author} {\bibfnamefont {A.~M.}\ \bibnamefont
  {Froese}}, \bibinfo {author} {\bibfnamefont {A.~I.}\ \bibnamefont
  {Smolyakov}}, \ and\ \bibinfo {author} {\bibfnamefont {D.}~\bibnamefont
  {Sydorenko}},\ }\bibfield  {title} {\enquote {\bibinfo {title} {Nonlinear
  skin effect in a collisionless plasma},}\ }\href {\doibase 10.1063/1.3211196}
  {\bibfield  {journal} {\bibinfo  {journal} {Physics of Plasmas}\ }\textbf
  {\bibinfo {volume} {16}},\ \bibinfo {pages} {080704} (\bibinfo {year}
  {2009})},\ \Eprint {http://arxiv.org/abs/https://doi.org/10.1063/1.3211196}
  {https://doi.org/10.1063/1.3211196} \BibitemShut {NoStop}%
\bibitem [{\citenamefont {Briefi}\ \emph {et~al.}(2022)\citenamefont {Briefi},
  \citenamefont {Zielke}, \citenamefont {Rauner},\ and\ \citenamefont
  {Fantz}}]{Briefi_2022}%
  \BibitemOpen
  \bibfield  {author} {\bibinfo {author} {\bibfnamefont {S.}~\bibnamefont
  {Briefi}}, \bibinfo {author} {\bibfnamefont {D.}~\bibnamefont {Zielke}},
  \bibinfo {author} {\bibfnamefont {D.}~\bibnamefont {Rauner}}, \ and\ \bibinfo
  {author} {\bibfnamefont {U.}~\bibnamefont {Fantz}},\ }\bibfield  {title}
  {\enquote {\bibinfo {title} {{Diagnostics of RF coupling in H$^-$ ion sources
  as a tool for optimizing source design and operational parameters}},}\ }\href
  {\doibase 10.1063/5.0077934} {\bibfield  {journal} {\bibinfo  {journal}
  {Review of Scientific Instruments}\ }\textbf {\bibinfo {volume} {93}},\
  \bibinfo {pages} {023501} (\bibinfo {year} {2022})},\ \Eprint
  {http://arxiv.org/abs/https://doi.org/10.1063/5.0077934}
  {https://doi.org/10.1063/5.0077934} \BibitemShut {NoStop}%
\bibitem [{\citenamefont {Fantz}\ \emph {et~al.}(2018)\citenamefont {Fantz},
  \citenamefont {Hopf}, \citenamefont {Friedl}, \citenamefont {Cristofaro},
  \citenamefont {Heinemann}, \citenamefont {Lishev},\ and\ \citenamefont
  {Mimo}}]{Fantz_2018}%
  \BibitemOpen
  \bibfield  {author} {\bibinfo {author} {\bibfnamefont {U.}~\bibnamefont
  {Fantz}}, \bibinfo {author} {\bibfnamefont {C.}~\bibnamefont {Hopf}},
  \bibinfo {author} {\bibfnamefont {R.}~\bibnamefont {Friedl}}, \bibinfo
  {author} {\bibfnamefont {S.}~\bibnamefont {Cristofaro}}, \bibinfo {author}
  {\bibfnamefont {B.}~\bibnamefont {Heinemann}}, \bibinfo {author}
  {\bibfnamefont {S.}~\bibnamefont {Lishev}}, \ and\ \bibinfo {author}
  {\bibfnamefont {A.}~\bibnamefont {Mimo}},\ }\bibfield  {title} {\enquote
  {\bibinfo {title} {{Technology developments for a beam source of an NNBI
  system for DEMO}},}\ }\href {\doibase
  https://doi.org/10.1016/j.fusengdes.2018.02.025} {\bibfield  {journal}
  {\bibinfo  {journal} {Fusion Engineering and Design}\ }\textbf {\bibinfo
  {volume} {136}},\ \bibinfo {pages} {340--344} (\bibinfo {year} {2018})},\
  \bibinfo {note} {special Issue: Proceedings of the 13th International
  Symposium on Fusion Nuclear Technology (ISFNT-13)}\BibitemShut {NoStop}%
\bibitem [{\citenamefont {Tran}\ \emph {et~al.}(2022)\citenamefont {Tran},
  \citenamefont {Agostinetti}, \citenamefont {Aiello}, \citenamefont
  {Avramidis}, \citenamefont {Baiocchi}, \citenamefont {Barbisan},
  \citenamefont {Bobkov}, \citenamefont {Briefi}, \citenamefont {Bruschi},
  \citenamefont {Chavan} \emph {et~al.}}]{Tran_2022}%
  \BibitemOpen
  \bibfield  {author} {\bibinfo {author} {\bibfnamefont {M.}~\bibnamefont
  {Tran}}, \bibinfo {author} {\bibfnamefont {P.}~\bibnamefont {Agostinetti}},
  \bibinfo {author} {\bibfnamefont {G.}~\bibnamefont {Aiello}}, \bibinfo
  {author} {\bibfnamefont {K.}~\bibnamefont {Avramidis}}, \bibinfo {author}
  {\bibfnamefont {B.}~\bibnamefont {Baiocchi}}, \bibinfo {author}
  {\bibfnamefont {M.}~\bibnamefont {Barbisan}}, \bibinfo {author}
  {\bibfnamefont {V.}~\bibnamefont {Bobkov}}, \bibinfo {author} {\bibfnamefont
  {S.}~\bibnamefont {Briefi}}, \bibinfo {author} {\bibfnamefont
  {A.}~\bibnamefont {Bruschi}}, \bibinfo {author} {\bibfnamefont
  {R.}~\bibnamefont {Chavan}},  \emph {et~al.},\ }\bibfield  {title} {\enquote
  {\bibinfo {title} {{Status and future development of Heating and Current
  Drive for the EU DEMO}},}\ }\href {\doibase
  https://doi.org/10.1016/j.fusengdes.2022.113159} {\bibfield  {journal}
  {\bibinfo  {journal} {Fusion Engineering and Design}\ }\textbf {\bibinfo
  {volume} {180}},\ \bibinfo {pages} {113159} (\bibinfo {year}
  {2022})}\BibitemShut {NoStop}%
\end{thebibliography}%
		
\end{document}